\documentclass[pre,nofootinbib,amsmath,amssymb,preprint,showpacs]{revtex4}
\usepackage[T1]{fontenc} \usepackage[latin1]{inputenc}
\usepackage{amsmath} \usepackage{graphicx} \usepackage{amssymb}
\begin{document}

\title{Structural modes of a polymer in the repton model}

\author{Gerard T. Barkema$^{\dagger\ddagger}$}
\affiliation{$^{\dagger} $Institute for Theoretical Physics,
Universiteit Utrecht,  Leuvenlaan 4, 3584 CE Utrecht, The
Netherlands\\ $^{\ddagger}$Instituut-Lorentz, Universiteit Leiden,
Niels Bohrweg 2, 2333 CA Leiden, The Netherlands} \author{Debabrata
Panja} \affiliation{Institute for Theoretical Physics, Universiteit
van Amsterdam, Valckenierstraat 65, 1018 XE Amsterdam, The
Netherlands} \author{J.M.J. van Leeuwen} \affiliation{
Instituut-Lorentz, Universiteit Leiden, Niels Bohrweg 2, 2333 CA
Leiden, The Netherlands}
\begin{abstract} 
  Using extensive computer simulations, the behavior of the structural
  modes --- more precisely, the eigenmodes of a phantom Rouse polymer
  --- are characterized for a polymer in the three-dimensional repton
  model, and are used to study the polymer's dynamics at time scales
  well before the tube renewal. Although these modes are not the
  eigenmodes for a polymer in the repton model, we show that
  numerically the modes maintain a high degree of statistical
  independence. The correlations in the mode amplitudes decay
  exponentially with $(p/N)^2A(t)$, in which $p$ is the mode number,
  $N$ is the polymer length and $A(t)$ is a single function shared by
  all modes. In time, the quantity $A(t)$ causes an exponential decay
  for the mode amplitude correlation functions for times $<1$; a
  stretched exponential with an exponent $1/2$ between times $1$ and
  $\tau_R\sim N^2$, the time-scale for diffusion of tagged reptons
  along the contour of the polymer; and again an exponential decay for
  times $t>\tau_R$.  Having assumed statistical independence and the
  validity of a single function $A(t)$ for all modes, we compute the
  temporal behavior of three structural quantities: the vectorial
  distance between the positions of the middle monomer and the
  center-of-mass, the end-to-end vector, and the vector connecting two
  nearby reptons around the middle of the polymer.  Furthermore, we
  study the mean-squared displacement of the center-of-mass and the
  middle repton, and their relation with the temporal behavior of the
  modes.
\end{abstract}

\pacs{36.20.-r,64.70.km,82.35.Lr}

\maketitle

\section{Introduction: polymer reptation and the repton
  model\label{sec:intro}}

\begin{figure}[h]
\includegraphics[width=0.5\linewidth]{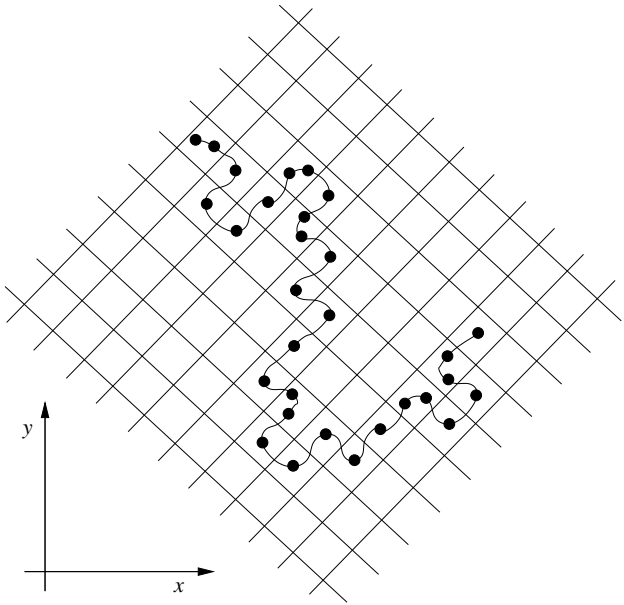}
\caption{A two-dimensional lattice representation of the repton
  model. The polymer is represented by points (reptons) that reside in
  the cells of the lattice, and are connected by bonds. The rules
  governing this model are described in the text.\label{fig:repton}}
\end{figure}

The dynamics of a coarse-grained polymer has two extremes in polymer
physics. One extreme features the phantom Rouse model \cite{de,doi},
wherein the polymer is described as a chain of beads serially
connected by harmonic springs. A polymer in the phantom Rouse model
does not feel its surroundings, and not even itself. Comprised of
$N+1$ beads with spatial locations $\vec r_i$ ($i=0,1,\ldots, N$), a
polymer of length $N$ in the Rouse model lends itself to a complete
analysis of its dynamics in terms of the eigenmodes indexed by $p$
($p=0,1,\ldots, N$), correspondingly called the Rouse modes, whose
amplitudes are given by
\begin{equation}
  \vec{X}_p(t)=\frac{1}{N+1}\sum_{i=0}^N\cos\left[\alpha_p(i+1/2)\right]
  \vec{r}_i (t),
\label{eq:rousemodes}
\end{equation}
in which $\alpha_p=p\pi/(N+1)$.  The other extreme features
Rubinstein's repton model~\cite{rubinstein}, which describes the
dynamics of a highly restricted polymer, in which the only dynamical
freedom is a mechanism known as \emph{reptation}~\cite{degennes,doi},
a snake-like motion in which stored length travels back and forth
along the polymer, and can be created and annealed only at the
polymer's ends.  In the repton model, the polymer is composed of
mobile points, called reptons, residing in the cells of a lattice,
connected by bonds.  A configuration of the polymer in the repton
model in two dimensions is shown in Fig. \ref{fig:repton}. The
neighboring reptons can reside either in the same cell or in the
adjacent cells. The configuration of the polymer's contour in space,
defined by the serially connected cells occupied by the reptons, is
that of a random walk, and is not allowed to undergo any change, other
than by the motion of one of its ends.  Each repton, except the end
ones, is therefore constrained to move along the contour of the
polymer. The repton model is extensively discussed and illustrated in
the literature \cite{widom,vanl}. Duke has extended the repton model
to study gel electrophoresis~\cite{duke}, by adding an electric field
which acts on the charged reptons; also this model has been studied
extensively~\cite{vanl,Drepton,tolya}. A highly efficient simulation
approach of the repton model was developed based on
bit-operations~\cite{Drepton}; the present work uses this approach as
well.

In this paper we characterize the analogs of the Rouse modes for a
polymer consisted of $N+1$ reptons in the repton model in three
spatial dimensions, having denoted the spatial locations of the
reptons by $\vec r_i$ ($i=0,1,\ldots N$), as in
Eq. (\ref{eq:rousemodes}). Henceforth, these are simply referred to as
the \emph{modes}.  They are not the eigenmodes of the polymer in the
repton model. We express several dynamical variables for the polymer
in terms of the modes. We verify the time-correlations of
the modes and these dynamical variables by high-precision
computer simulations.

In the simulations the repton positions are the natural variables. For
theoretical analysis however, the distance between successive reptons,
or the ``link variables'', defined as
\begin{equation}
\vec{y}_i = \vec{r}_i - \vec{r}_{i-1} \quad \quad \quad {\rm for} \quad i=0, \cdots , N.
\label{eq:y}
\end{equation} 
are a more convenient choice. By construction, the link variables are
either a unit vector, or zero. In the former case, the link is taut,
while in the latter case the link is slack and it represents an
element of ``stored length''. The motion of the polymer can be viewed
as diffusion of stored length along the chain (instead of the
movements of the individual reptons).

Few exact results are known for the repton model, all concerning the
asymptotic time regime, governed by the overall diffusion constant $D$
of the polymer. It was already proposed by
Rubinstein~\cite{rubinstein}, that $D$ scales as $N^{-2}$.  Later it
was found by van Leeuwen and Kooiman \cite{kooiman} that in the repton
model on a $d$-dimensional hypercubic lattice, $D=1/[(2d+1)N^2]$.
This result was shown to be exact by Pr\"ahofer and Spohn
\cite{prahofer} and by Widom and Al-Lehyani \cite{mwidom}. In this
paper, we study the repton model on a cubic lattice. For reasons of
computational efficiency, the rates for extraction and retraction of
the ends are modified to obtain a dimensionality-independent density
of stored length, tuned to that of the one-dimensional model, so that
$D=1/(3N^2)$. We are specifically interested in the polymer's dynamics
over intermediate time regimes, for which, so far, only simulations
can reveal the behavior.

Before the diffusive regime is reached, the repton model has various
distinct time regimes. First, at very short times $t<1$, the dynamics
is governed by individual hops of reptons, which are spatially
separated.  Next, there is one intermediate regime $1<t<\tau_R\sim
N^2$, where $\tau_R$ is the time at which the fluctuations in the
density of stored lengths decay. This is followed by a second
intermediate regime $\tau_R<t<\tau_d\sim N^3$, where $\tau_d$ is the
well-known tube-renewal time after which the initial information on
the chain has been forgotten.

Our basic quantity of interest is the equilibrium correlation function
\begin{equation}
C_{pq} (t) \equiv \left< \vec{X}_p(t) \cdot \vec{X}_q(0)\right>,
\label{eq:corr}
\end{equation}
in the above time regimes, where the angular brackets denote an
average over the equilibrium ensemble of polymer conformations. More
specifically, we put the middle monomer at $t=0$ at the origin. Then a
polymer configuration can be build up from independent link variables
$\vec y_i$. The link may be taut or slack, with probabilities $2/3$
and $1/3$. A taut link variable may assume one of the $6$
possibilities corresponding to the six directions on the cubic
lattice, each with equal weight.

Since the modes are not eigenmodes of the polymer in the repton model,
they do not decay exponentially in time. In Sec. \ref{sec:cross} we
show that the cross-correlation functions for $p \neq q$ vanish for
$t=0$, and remain small at later times. In Sec. \ref{sec:auto} we
study the autocorrelation functions $C_{pp}(t)$. These
autocorrelations are found to decay exponentially at short times $t<1$
and at long times $t>\tau_R$, but in the intermediate regime
$1<t<\tau_R$ they decay as stretched exponentials in time, with
exponent $1/2$. In all these regimes, the exponent has a power-law
dependence on mode number $p$ and polymer length $N$. In
Sec. \ref{sec:struc}, using the mode amplitude autocorrelation
functions $C_{pp}(t)$ and neglecting cross-correlations, we
reconstruct other time-dependent correlation functions, in particular
those of the vector between the positions of the middle repton and the
center-of-mass, the end-to-end vector, and a spatial vector connecting
two nearby reptons around the middle repton. We compare these analytic
results with direct computer simulations.  In Sec. \ref{sec5} we
evaluate the mean-squared displacement of the center-of-mass and of
the middle repton in the above time regimes, and compare them with
direct computer simulations. We end the paper with a discussion in
Sec. \ref{sec:discussion}.

\section{Behavior of correlations between different modes
\label{sec:cross}}

For phantom chains in the Rouse model, the modes as defined
in Eq.~(\ref{eq:rousemodes}) are strictly independent and orthogonal, i.e.
the correlations $C_{pq}(t)$ as defined in Eq.~(\ref{eq:corr})
are zero for all times, if $p\neq q$. This property is part of the
motivation why in the study of the dynamics of a polymer, it is very
convenient to express the quantities of interest in amplitudes of the
structural modes.  In this section we will show analytically that at
$t=0$, also in the repton model the same structural modes are orthogonal.
We will also show that correlations between even modes $p$ and odd modes
$q$ are strictly zero at all times.  Next, with computer simulations, we
will show that for time $t>0$ cross-correlations between the modes $p\neq
q$, in which $p$ and $q$ are both even or both odd, start to develop.

\subsection{Behavior of $C_{pq}(0)$}

First, we turn to the correlation functions $C_{pq}(t)$ for
time $t=0$, which only require the evaluation of equilibrium averages.
For this purpose it is convenient to provide the relation between the
mode amplitudes and the link variables which have the simple
equilibrium averages
\begin{equation} 
\langle \vec y_j \,\vec y_k \rangle = {\bf  I}\, (\rho/d) \,
\delta_{jk},
\label{a4} 
\end{equation} 
where ${\mathbf I}$ is the unity tensor, and $\rho$ is the equilibrium
density of the taut links, which equals $\rho=2/3$ in our simulations.

We express the position of the $i$-th repton w.r.t. that of the
middle repton $\vec r_{N/2}$ [for simplicity we take $N$ even all
throughout this paper; if one is interested in odd $N$
polymers, one can take as a reference point $\vec r_m=\frac12\left(
\vec r_{(N-1)/2}+\vec r _{(N+1)/2}\right)$, which has the same
symmetry] and the link variables as 
\begin{equation} 
  \vec{r}_j = \vec{r}_{N/2} + \sum ^j
  _{i=N/2+1} \vec{y}_i, \quad {\rm for} \,(j>N/2), \quad \quad \quad
  \vec{r}_j = \vec{r}_{N/2} - \sum ^{N/2} _{i=j+1} \vec y_i \quad
  {\rm for} \, (j<N/2).
\label{b1} 
\end{equation} 
In the summations to come we will use the following relation for $p \neq 0$
\begin{equation} 
\sum^N_{j=k}\cos[\alpha_p (j+1/2)] = -{\frac{\sin ( \alpha_p k)}{2
    s_p}},
\label{b2}
\end{equation} 
with $s_p = \sin (\alpha_p/2)$.  Note that Eq. (\ref{b2}) yields 0 for
$k=0$. This implies that the middle repton location $\vec{r}_{N/2}$ in
Eq. (\ref{b1}) does not contribute in the mode amplitude expression
for $p \neq 0$.  Thus, using Eq. (\ref{b1}), and further, upon
interchanging the summations over $j$ and $k$, for $p \neq 0$, we have
\begin{equation}
\vec{X}_p = - \frac1{2 (N+1) s_p} \sum^N_{k=1} \, \sin (\alpha_p k)\,
\vec{y}_k .
\label{b3}
\end{equation}
Equation (\ref{b3}) cannot be applied to the center-of-mass mode $p=0$
as $s_0=0$.  For $p=0$, with the help of Eq. (\ref{b1}), the
corresponding formula becomes
\begin{equation} 
\vec{X}_0 = \vec{r}_{N/2} + \frac{1}{N+1}
  \left[\sum ^{N/2}_{j=1} j\,  (\vec{y}_{N+1-j} -   \vec{y}_j)
  \right].
  \label{b4} 
\end{equation} 
While the mode amplitudes for $p \neq 0$ are expressed solely
in terms of the link (or structural) variables, the center-of-mass
mode needs an additional (translational) coordinate $\vec{r}_{N/2}$.
 
Equation (\ref{b3}) can be inverted using the relation 
\begin{equation} 
\sum^N_{k=1} \sin ( \alpha_p k) \sin (\alpha_q k)= \frac {N+1}{2}\,
\delta_{pq}
\label{b6}
\end{equation} 
and its inverse
\begin{equation}
\sum^N_{p=1} \sin (\alpha_p k) \sin (\alpha_p l)= \frac {N+1}{2}\,
\delta_{kl}.
 \label{b7}
\end{equation}
These results lead us to the inverse relation for the link variables
in terms of the mode amplitudes
\begin{equation} 
\vec{y}_k = -4 \sum^N_{p=1} s_p \, \sin (\alpha_p k )\,\vec{X}_p.
\label{b10}
\end{equation} 
Equation (\ref{b10}) reveals that the mode amplitudes for $p \neq 0$
and the link variables are fully equivalent since they can be
transformed into each other.

From the expression (\ref{b4}) and the average (\ref{a4}) it follows
that the mode amplitudes for $p \neq 0$ and $q \neq 0$ are orthogonal
for $t=0$ with the inner product
\begin{equation}  
C_{pq} (0) = \frac{\rho }{8  (N+1) s^2_p } \delta_{pq}.
\label{b11}
\end{equation}  

\subsection{Behavior of $C_{pq}(t)$ for $p,q>0$ and $t>0$}

\begin{figure}
\begin{center}
\includegraphics[width=0.97\linewidth]{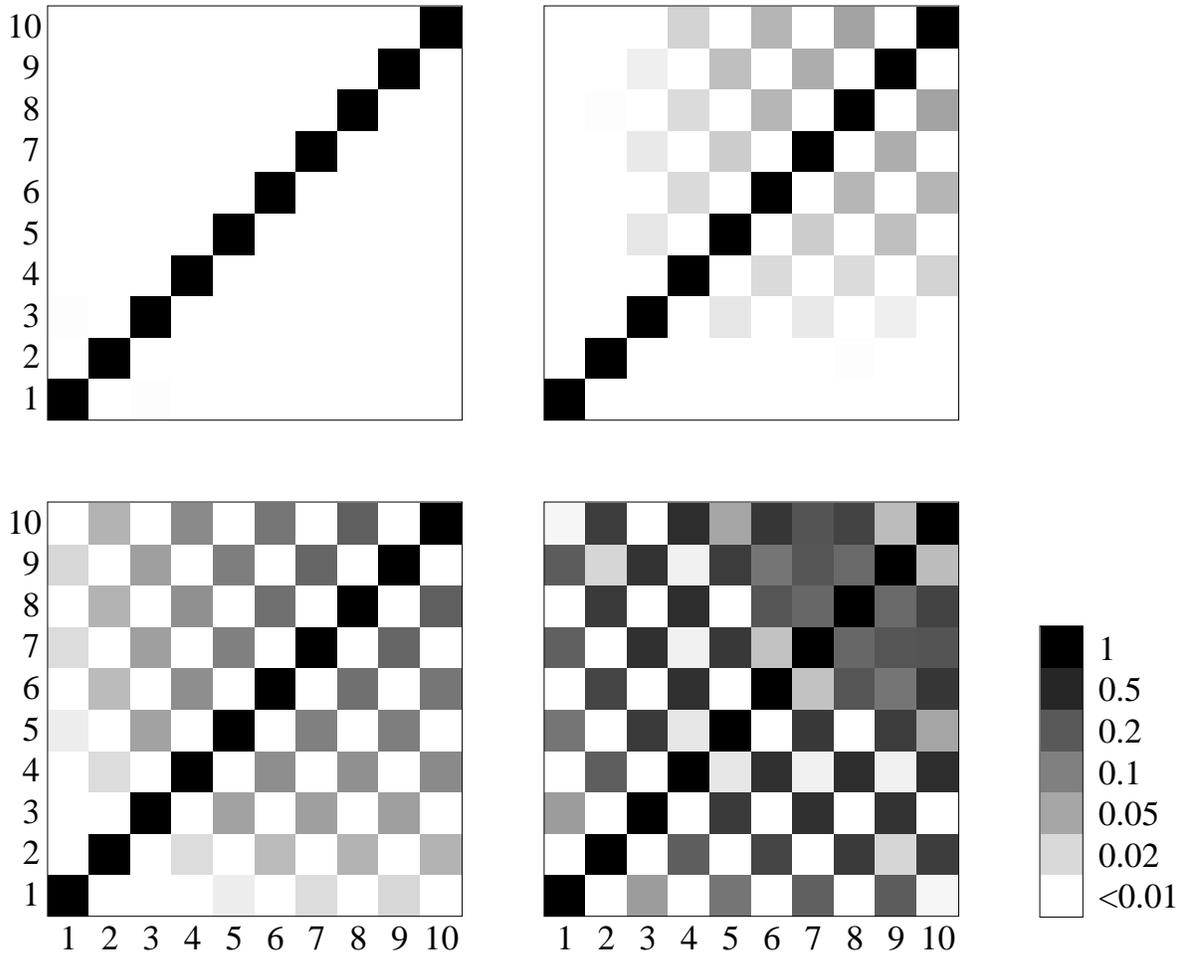}
\end{center}
\caption{Checkerboard plot for $\eta_{pq}(t)$ for $N=400$ at four
different times; upper left: $t=0$, upper right: $t=8,000$, lower
left: $t=64,000$, and lower right $t=10^6$. The data are manually
symmetrized around the diagonal. The grayscale used is shown in the
scale bar.}
\label{checker}
\end{figure}

In Fig. \ref{checker} we plot the normalized cross-correlation
\begin{equation} \label{c1}
\eta_{pq}(t)=C_{pq}(t)/\sqrt{C_{pp}(t)C_{qq}(t)}
\end{equation} 
for a number of $p$ and $q$ and several times.  In the upper left
panel one observes the absence of cross-correlations for $t=0$, as
derived in Eq.  (\ref{b11}). The next clear observation in
Fig. \ref{checker} is the absence of cross-correlations between the
even and odd modes. This is a consequence of the inversion symmetry of
the polymer chain: inversion is a renumbering of the reptons from $N$
to 0 and the inversion of the link variables, or notationally,
\begin{equation} \label{c2} \vec{y}_i \leftrightarrow
-\vec{y}_{N-i+1}.
\end{equation}  
The amplitudes of the even modes are invariant under this operation,
while the odd mode amplitudes change sign. Since the dynamical
evolution commutes with the inversion operation, the symmetry of the
even mode amplitudes and the change of sign for the odd mode
amplitudes are preserved under time evolution. In the lower-right
panel the inversion symmetry appears to be violated for larger $p$ and
$q$, but this is actually a consequence of the numerical inaccuracy of
the simulation data, which is magnified once the normalizations
involved in $\eta_{pq}$ become very small.

Figure \ref{checker} also reveals that for $t>0$, $C_{pq}(t)$ is no
longer strictly proportional to $\delta_{pq}$. Nevertheless, the
off-diagonal correlation functions are small in comparison to
the diagonal ones.

\section{Behavior of $C_{pp}(t)$\label{sec:auto}}

For the properties for $C_{pp}(t)$ we can anticipate the following
behavior.
\begin{itemize}
\item[(i)] At short times $t<1$, reptons typically make zero or one
  move, and the movements of individual reptons are uncorrelated. In
  an earlier work on lattice polymers with exactly the same equilibrium
  properties as in the repton model, but less restricted
  dynamics~\cite{rousemodes}, we found that the decay of the
  structural modes in this short-time regime was characterized by
  $C_{pp}(t)=C_{pp}(0) \exp\left[ -A_1 p^2t/N^2\right]$, for some
  constant $A_1$.  Since the moves allowed in the repton model are a
  finite fraction of the moves in this earlier work, we expect the
  same behavior to hold, be it with a modified value of the parameter
  $A_1$.
\item[(ii)] In the regime $t>1$ reptons have the time to attempt
  multiple moves.  Right after a repton has moved in the ``upstream''
  direction, the reverse ``downstream'' move is guaranteed to be
  possible, while another ``upstream'' move requires the presence of
  another slack link upstream, and therefore has a significant
  probability to be ruled out. Hence, the nature of the dynamics
  changes at $t \sim 1$. From a curvilinear perspective, the motion
  changes from ordinary diffusion to ``single-file diffusion'', in
  which the squared curvilinear displacement increases $\sim
  \sqrt{t}$. Continuity of the curves near $t\sim 1$ then leads to an
  expected behavior $C_{pp}(t)=C_{pp}(0) \exp\left[ -A_2
    p^2\sqrt{t}/N^2\right]$.
\item[(iii)] Slack links can move freely up and down the chain, in
  unit steps, without constraints.  This diffusive behavior along the
  chain with a curvilinear diffusion constant of $\sim 1$, combined
  with the polymer length $\sim N$, yields a characteristic time scale
  $\tau_R \sim N^2$ beyond which the distribution of slack links
  becomes statistically independent. Without such correlations, the
  curvilinear dynamics ceases to be anomalous and becomes ordinary
  diffusion once more. Continuity of all curves around $\tau_R$ then
  yields $C_{pp}(t)=C_{pp}(0) \exp\left[ -A_3 p^2t/N^3\right]$. The
  data reveal that $A_1$, $A_2$ and $A_3$ are all of $O(1)$.
\end{itemize}

\begin{figure*}
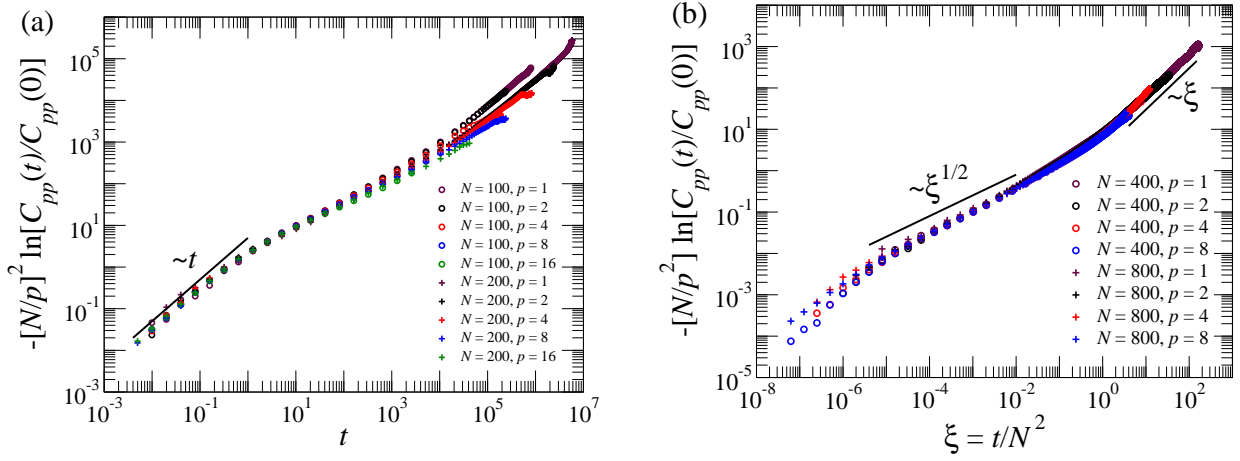

\begin{center}
\begin{minipage}{0.48\linewidth}
\includegraphics[width=\linewidth]{shorttime}
\end{minipage} \hspace{4mm}
\begin{minipage}{0.48\linewidth}
\includegraphics[width=0.97\linewidth]{longtime}
\end{minipage}
\end{center}
\caption{Behavior of $C_{pp}(t)$ for $t>0$. (a) for $t<1$,
$C_{pp}(t)=C_{pp}(0)\exp[-A_1p^2t/N^2]$ for some constant $A_1$ of
magnitude $O(1)$. (b) for $1<t<\tau_R\sim N^2$,
$C_{pp}(t)=C_{pp}(0)\exp[-A_2p^2t^{1/2}/N^2]$, while for $t>\tau_R$,
$C_{pp}(t)=C_{pp}(0)\exp[-A_3p^2t/N^3]$, with $A_2$ and $A_3$ also of magnitude
$O(1)$.}
\label{autocor}
\end{figure*}

Having combined (i-iii), we present the simulation data for several
$p$ and $N$ values in Fig. \ref{autocor}. The data collapse confirms that
for $t<1$
\begin{eqnarray} C_{pp}(t)=C_{pp}(0)\exp[-A_1p^2t/N^2]
\label{e9}
\end{eqnarray} 
for some constant $A_1$ of magnitude $O(1)$. Further, for
$1<t<\tau_R\sim N^2$
\begin{eqnarray} C_{pp}(t)=C_{pp}(0)\exp[-A_2p^2t^{1/2}/N^2],
\label{e10}
\end{eqnarray} while for $t>\tau_R$
\begin{eqnarray} C_{pp}(t)=C_{pp}(0)\exp[-A_3p^2t/N^3].
\label{e11}
\end{eqnarray} 
Having combined Eqs.~(\ref{e9}-\ref{e11}), we see that $C_{pp}(t)$
is expressed in general terms as $C_{pp}(0)\exp[-(p/N)^2A(t)]$ with a
time-dependent quantity $A(t)$; which we will encounter in the coming
sections. At short and long times, this function $A(t)$ is linear in time,
with at short times a length-independent prefactor $A_1$ and at long times
a prefactor $A_3/N$, inversely proportional to polymer length;
in the intermediate regime $1<t<\tau_R \sim N^2$, the function $A(t)$
increases proportional to $\sqrt{t}$, with a length-independent prefactor
$A_2$.  Note that at all times $A(t)$ is independent of the mode number $p$.

\section{Three structural quantities in terms of mode amplitudes\label{sec:struc}}

All structural quantities which can be expressed in terms of the link
variables, are also expressible in terms of the mode amplitudes. In the
previous section, we obtained an approximate description of the behavior
in time of the mode amplitudes: we neglect cross-correlations at all times,
and assume that a single function $A(t)$ determines the autocorrelation 
of all mode amplitudes. In this section, we will confront this approximate
description with direct simulations. For this purpose, we
consider three such structural variables: the vector
$\vec{S}_0$ between the center-of-mass and the middle repton, the
polymer's end-to-end vector $\vec S_1$, and the vector $\vec S_2$
connecting the two reptons next to the middle repton.

\subsection{Structural variables $S_0$, $S_1$ and $S_2$ expressed in
  terms of the modes amplitudes\label{strucsec}}

The distance $\vec{S}_0$ between the center-of-mass and the middle
repton, defined in Eq. (\ref{b4}), is given by
\begin{equation} 
  \vec{S}_0 = \vec{R}_0 -\vec{r}_{N/2} = \frac{1}{N+1}
  \left[\sum ^{N/2}_{j=1} j\,  (\vec{y}_{N+1-j} -   \vec{y}_j)
  \right],
  \label{b12}
\end{equation} 
in terms of the link variables. Its mean-squared equilibrium-average
follows from Eq.~(\ref{a4}) as
\begin{equation} 
\langle S^2_0 \rangle = \frac{\rho N(N+2)}{12 (N+1)}.
\label{b13}
\end{equation} 
Having expressed the link variables in terms of the mode amplitudes
further yields the expression
\begin{equation} \label{b14}
\vec{S}_0 = 2 \sum^N_{p=2, 4, \cdots} (-1)^{p/2} \vec{X}_p.
\end{equation} 

Similarly, the end-to-end vector $\vec{S}_1$ of the polymer is
expressed as
\begin{equation} 
\vec{S}_1=\vec{r}_N - \vec{r}_0 = \sum^N_{k=1} \vec{y}_k.
\label{b15}
\end{equation}  
in terms of the link variables. Its mean-squared equilibrium-average
is given by
\begin{equation} 
\langle S^2_1 \rangle = \rho N,
\label{b16} 
\end{equation}
and the expression of $\vec{S}_1$ in terms of the mode amplitudes
takes the form
\begin{equation} 
\vec{S}_1 = -4\sum^N_{p=1,3,\cdots}  c_p \, \vec{X}_p.
\label{b17} 
\end{equation} 
with $c_p=\cos(\alpha_p /2)$. 

Note that the end-to-end vector is a special case of the vector
between two arbitrary reptons $m$ and $n$
\begin{equation} 
\vec{r}_m - \vec{r}_n = \sum^m_{k=n+1}\vec{y}_k.
\label{b18} 
\end{equation}
For a characteristic case, we consider the vector $\vec{S}_2$
between the two reptons next to the middle repton as our third
structural variable, i.e.,
\begin{equation} 
\vec{S}_2 = \vec{r}_{N/2+1} -\vec{r}_{N/2-1},
\label{b19}
\end{equation}  
with the mean-squared equilibrium-average
\begin{equation} 
\langle S^2_2 \rangle = 2 \rho;
\label{b20} 
\end{equation}
$\vec S_2$ is related to the mode amplitudes as
\begin{equation} 
\vec{r}_{N/2+1} -\vec{r}_{N/2-1} =  -8 \sum^N_{p=1} \sin(p \pi/2)\,
c_p s_p \vec{X}_p.
\label{b21}
\end{equation} 

We summarize the expressions of $\vec S_0$, $\vec S_1$ and $\vec S_2$
[Eqs. (\ref{b14}), (\ref{b17}) and (\ref{b21})] in the form
\begin{equation} 
\vec{S}_i = \sum^N_{p=1} S_{p,i} \vec{X}_p.
\label{b22}
\end{equation} 
The equilibrium averages can also be evaluated via the averages of the
mode amplitudes, as provided in Eq. (\ref{b11}). This leads one to the
evaluation of sums of the type
\begin{equation} 
\langle S_i^2 \rangle = \frac{\rho }{8  (N+1)}\sum^N_{p=1} (S_{p,i}/s_p)^2,
\label{b23}
\end{equation} 
which can be found in Ref. \cite{book}.

Although the mode amplitudes for $p \neq 0$ are equivalent to the link
variables, one must realize that the domains of possible values are
quite different. The domain of possible $\vec y_j$ consists of
$2d+1=7$ values for each $j$ in three dimensions $(d=3)$, whereas the
domain of possible values of the $\vec{X}_p$ is complicated.  Each
$\vec{X}_p$ assumes $(2d+1)^N$ values (multiplicities included) and
the possibilities are inter-related. 

\subsection{Dynamical properties of the structural variables $\vec{S}$\label{structural}} 

Whereas the equilibrium averages of structural properties are easily
evaluated using the link variables, their time-dependent correlators
have to be evaluated through the use of temporal behavior of the
correlations in the mode amplitudes.

Using the general form Eq.~(\ref{b22}), we find, for the correlators,
that
\begin{equation} 
  \langle \vec S_i (t) \cdot  \vec S_i (0) \rangle = \sum^N_{p,q=1} S_{p,i}
  S_{q,i} \, C_{pq} (t).
\label{d1} 
\end{equation} 
For the time dependent $C_{pq} (t)$ we use the approximation discussed
in the previous section, leaving out the cross-correlations and
employing for the diagonal terms the generic form
\begin{equation} 
C_{pp} (t) = C_{pp} (0) \exp[ -(p/N) ^2 A(t)] =
 \frac{\rho} {8 (N+1) s^2_p} \exp[ -(p/N) ^2 A(t)].
\label{d2} 
\end{equation} 
In other words,
\begin{equation} 
 \langle \vec S_i (t) \cdot  \vec S_i (0) \rangle = 
\frac{8 \rho}{N+1} \sum^N_{p=1} (S_{p,i}/s_p)^2 \exp[ -(p/N)^2 A(t)].
\label{d3}
\end{equation} 
We note that the $t=0$ values are given as sums over the ratios
$(S_{p,i}/s_p)^2$, which are most easily evaluated through their
relation with the link variable representation.

For the finite chains that we consider here, numerical evaluation of
the sums presents no problems. The behavior can be generally
characterized by three regimes.

\begin{enumerate}
\item The initial regime, in which $A(t)$ remains of order unity and
  all terms in the sum over $p$ participate. In this regime the
  deviation from the initial value is proportional to $A(t)$.
\item The intermediate regime from $1 \leq t \leq \tau_R$ where $A(t)$
  grows as $\sqrt{t}$ with time. Then the small $p$ modes contribute
  to the sum, but the exponential effectively cuts the sum off at a
  value of $p \sim N/\sqrt{A(t)}$.
\item The asymptotic regime, starting from the Rouse time $\tau_R$.
  Then $A(t)$ is of order $1/N$ and the sum is
  effectively restricted to its first term with exponentially small
  corrections for the other $p$.
\end{enumerate}

In order to check the approximation (\ref{d3}) we performed computer
simulations for chains with length $N=400$, and stored at regular
times the instantaneous values for the structural variables
$\vec{S}_0$, $\vec{S}_1$ and $\vec{S}_2$. These observables show a
dynamical behavior with variations spanning over many decades. For
practical purposes, we therefore performed independent simulations
with $n_{\Delta t}=1, 10^2, 10^4, 10^6$ or $10^7$ elementary moves
between successive measurements, corresponding to time steps of
$\Delta t=n_{\Delta t}/(N+1)$. For each time step, we generated 64
sets of $10^5$ measurements. The results of these independent
simulations were then combined into a single curve in Fig. \ref{fig3}
(see also the next paragraph), which occasionally shows little
``hiccups'' at the times where the results of independent simulations
are joined together.

\vspace{7mm}
\begin{figure}[h]
\begin{center}
\includegraphics[width=0.6\linewidth]{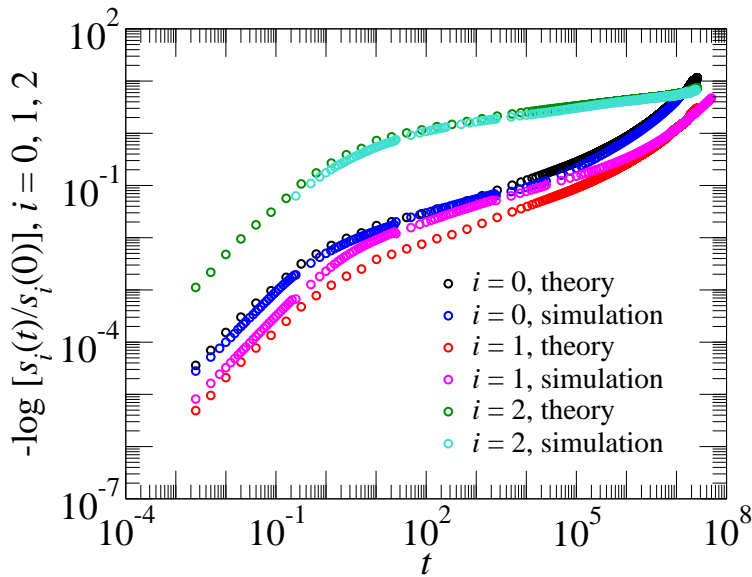}
\end{center}
\caption{The correlation functions $s_i (t) =\langle \vec{S}_i (t)
  \cdot \vec{S}_i (0) \rangle/ \langle S^2_i \rangle$ for $i=0$, $1$
  and $2$.}
\label{together}
\end{figure}
\begin{figure*}
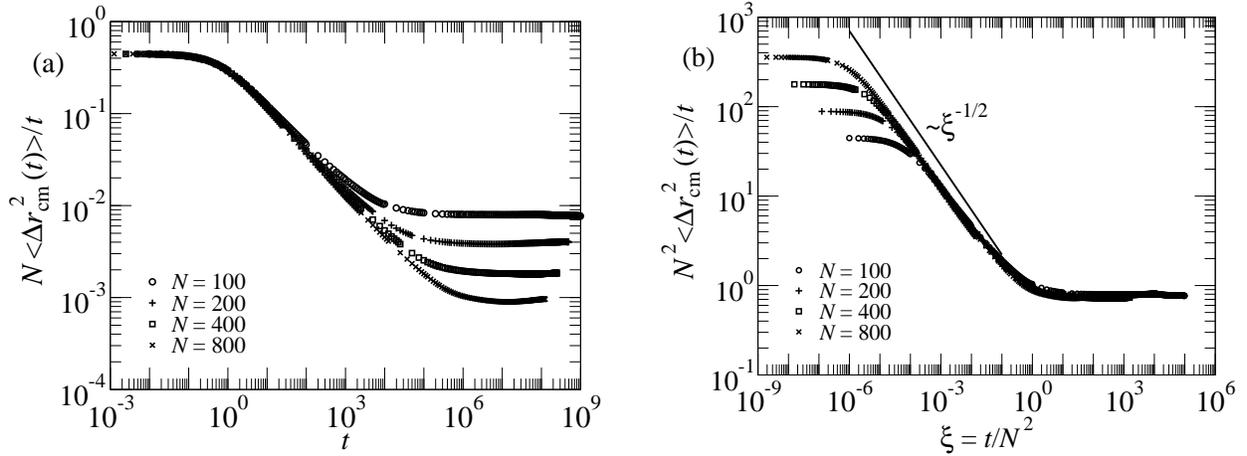

\begin{center}
\vspace{7mm}
\begin{minipage}{0.48\linewidth}
\includegraphics[width=\linewidth]{cmshorttime}
\end{minipage}
\hspace{4mm}
\begin{minipage}{0.48\linewidth}
\includegraphics[width=0.97\linewidth]{cmlongtime}
\end{minipage}
\end{center}
\caption{Behavior of the mean-squared displacement of the
  center-of-mass, $\langle\Delta r^2_{\text{cm}}(t)\rangle$. (a) for
  $t<1$, $\langle\Delta r^2_{\text{cm}}(t)\rangle\sim t/N$. (b) for
  $1<t<\tau_R\sim N^2$, $\langle\Delta r^2_{\text{cm}}(t)\rangle\sim
  \sqrt{t}/N$, while for $t>\tau_R$, $\langle\Delta
  r^2_{\text{cm}}(t)\rangle\sim t/N^2$.}
\label{fig3}
\end{figure*}

In Fig. \ref{together} we plot the measured value of the correlators
for $N=400$ and the calculated values from Eq.~(\ref{d3}). The
agreement is very good for $i=0$ and $i=2$; for $i=1$ the calculated
values are about a factor of two smaller than the measured values in
the intermediate range.  Given the very wide range in time and values
of these correlators the correspondence is still impressive. Since the
main approximation is the neglection of the cross-correlations in the
structural modes, this is indirect proof that although the number of
cross-correlations is very large, the main contribution comes from the
autocorrelations.

\section{Mean-squared displacement of center-of-mass and 
 middle repton\label{sec5}}

Besides the dynamic behavior of structural properties of the polymer,
also the dynamical behavior of its position is of interest. Two key
observables that characterize this, are the mean-squared displacement
of the center-of-mass, and of the middle repton.

We performed computer simulations of polymers with length $N=100, 200, 400$ and $800$,
and stored at regular times the positions of the middle repton and the center-of-mass.
For each length, we performed independent simulations in which the time between
updates of the positions was $n_{\Delta t}=1, 10^2, 10^4, 10^6$ or $10^7$ elementary
moves, corresponding to time differences of $\Delta t = n_{\Delta t}/N$. For each
length and time difference, we performed 64 simulations of $10^5$ time steps each.

We first show the results for the mean-squared displacement of the
center-of-mass and the middle repton of the polymer, and then we
discuss how the two are related.
\begin{figure*}
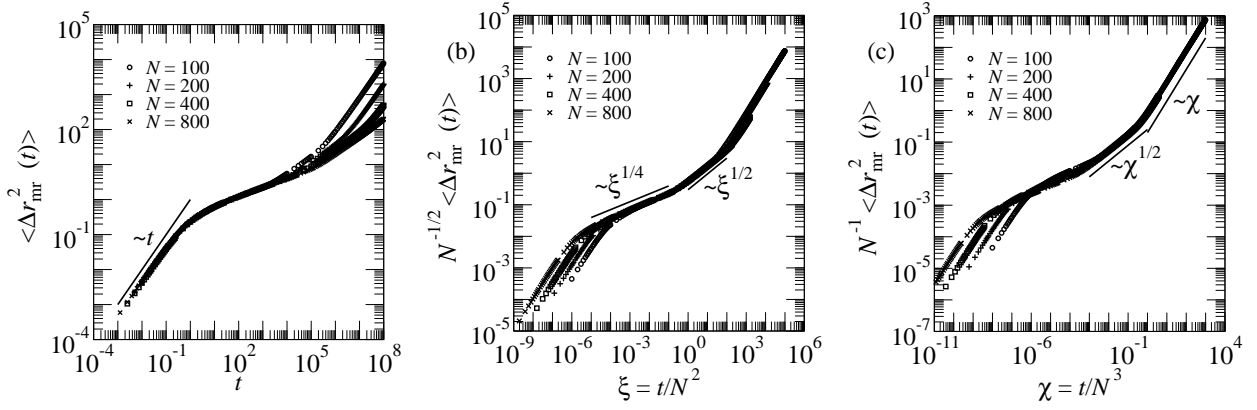

\begin{center}
\begin{minipage}{0.32\linewidth}
\includegraphics[width=\linewidth]{mmshorttime}
\end{minipage}
\hspace{1mm}
\begin{minipage}{0.32\linewidth}
\includegraphics[width=\linewidth]{mmmid}
\end{minipage}
\hspace{1mm}
\begin{minipage}{0.32\linewidth}
\includegraphics[width=\linewidth]{mmlong}
\end{minipage}
\end{center}
\caption{Behavior of the mean-squared displacement of the middle
  repton, $\langle\Delta r^2_{N/2}(t)\rangle$. (a) for $t<1$,
  $\langle\Delta r^2_{N/2}(t)\rangle\sim t$. (b) for
  $1<t<\tau_R\sim N^2$, $\langle\Delta r^2_{N/2}(t)\rangle\sim
  t^{1/4}$. (c) for $\tau_R<t<\tau_c$, $\langle\Delta
  r^2_{N/2}(t)\rangle\sim\sqrt{t/N}$, and for $t>\tau_c$,
  $\langle\Delta r^2_{N/2}(t)\rangle\sim t/N^2$.}
\label{fig4}
\end{figure*}

The computer simulations show that the mean-squared displacement
of the center-of-mass of the polymer behaves as follows:
\begin{eqnarray}
\langle[\vec{X}_0(t)-\vec{X}_0 (0)]^2\rangle\equiv
\langle \Delta^2_{\text{cm}} (t)\rangle\sim\left\{
\begin{array}{lll}
t/N&\quad t\lesssim1\\
t^{1/2}/N&\quad1\lesssim t\lesssim\tau_R\sim N^2\\
t/N^2&\quad t\gtrsim\tau_R
\end{array}\right..
\label{e15}
\end{eqnarray}
The data corresponding to Eq. (\ref{e15}) are shown in
Fig. \ref{fig3}. 

For the mean-squared displacement of the middle repton we find from the
simulations
\begin{eqnarray}
\langle [\vec{r}_{N/2} (t) - \vec{r}_{N/2} (0)]^2 \rangle \equiv 
\langle \Delta^2_{N/2}(t)\rangle\sim\left\{
\begin{array}{llll}
t&\quad t\lesssim1\\
t^{1/4}&\quad1\lesssim t\lesssim\tau_R\sim N^2\\
\sqrt{t/N}&\quad\tau_R\lesssim1\lesssim\tau_d \sim N^3\\
t/N^2&\quad t>\tau_d
\end{array}\right..
\label{e18}
\end{eqnarray}
The data corresponding to Eq. (\ref{e18}) are shown in
Fig. \ref{fig4}. Note that the microscopic formulation for the mean
squared displacement of the middle monomer is that of a Generalized
Langevin Equation \cite{panja_jstat}.

The two mean-squared displacements are related in the following manner
\begin{equation} 
\langle \Delta^2_{\text{cm}} (t)\rangle - \langle \Delta^2_{N/2}(t)\rangle =
2 \langle \vec{r}_ {N/2}(t) \cdot [\vec{S}_0 (t) - \vec{S}_0 (0)] \rangle 
+ \langle [\vec{S}_0 (t) - \vec{S}_0 (0)]^2 \rangle,
\label{e1}
\end{equation}
since $\vec{S}_0$ is the vectorial distance between the center-of-mass and the
middle repton.  The last term in Eq. (\ref{e1}) can be related to a
structural correlation function using the relation
\begin{equation} \label{e2a}
\langle S^2_0 (t) \rangle = \langle S^2_0 (0) \rangle. 
\end{equation} 

This equality holds because the structural distribution is not
affected by the position of the middle repton at $t=0$. So we may
evaluate this terms as
\begin{equation} \label{e3}
\langle [\vec{S}_0 (t) - \vec{S}_0 (0)]^2 \rangle =2[ \langle S^2_0 \rangle - 
\langle \vec{S}_0 (t) \cdot \vec{S}_0 (0) \rangle].
\end{equation} 
The correlator of $\vec{S}_0$ has been worked out in
Sec. \ref{structural} and the equilibrium value in Eq. (\ref{b13}).

\begin{figure}[t]
  \begin{center}
 \includegraphics[width=0.6\linewidth]{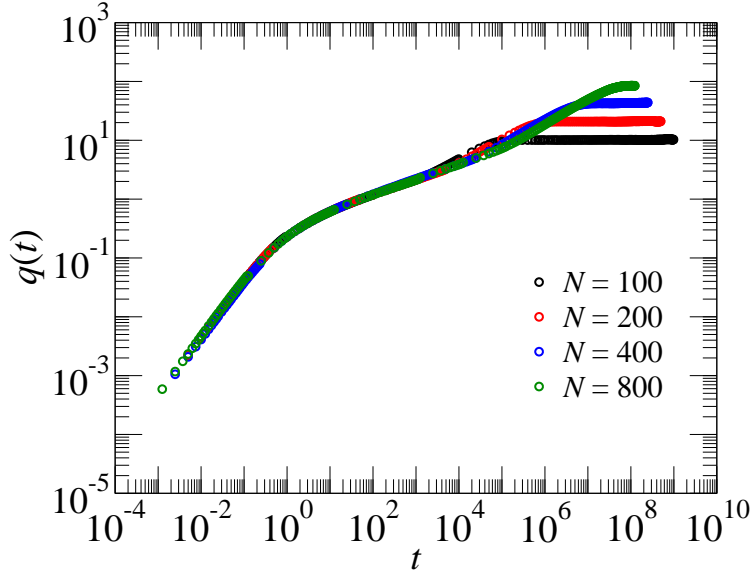}
 \end{center}
 \caption{The cross-correlation function $q(t)= -\langle
   \vec{r}_{N/2}(t) \cdot [\vec{S}_0 (t) - \vec{S}_0 (0)] \rangle$.} 
 \label{crosscor}
\end{figure}

The first term (\ref{e1}), being a cross-correlation between a
translational and a structural variable, cannot be computed using the
structural modes only. It is interesting to remark that
\begin{equation} 
\langle \vec{r}_{N/2}(t) \cdot [\vec{S}_0 (t) + \vec{S}_0 (0)]
\rangle =0,
\label{e4}
\end{equation} 
since the sum $\vec{S}_0 (t) + \vec{S}_0 (0)$ is time-reversal
invariant, while $\vec{r}_{N/2}(t) $ changes sign under time
reversal. We have verified that the simulations obey this consequence
of time-reversal invariance. We therefore have
\begin{equation} 
\langle \vec{r}_{N/2}(t) \cdot [\vec{S}_0 (t) - \vec{S}_0 (0)] \rangle=
-2 \langle \vec{r}_{N/2}(t) \cdot  \vec{S}_0 (0) \rangle.
\label{e5}
\end{equation}
It is clear that the right hand side of Eq. (\ref{e5}) is negative
because the middle repton has a tendency to move in the direction of
the instantaneous position of the center-of-mass. It is worth noting
that this tendency has a persistent effect on the value of the
correlation function. Similarly, it can be shown that the
center-of-mass also has a tendency to move towards the middle repton!
This means that a reptating polymer experiences more than only
internal forces --- indeed, part of the forces a reptating polymer
experiences originate from the dynamical constraints that prevent the
polymer from moving sideways; giving rise to the phenomenon that mode
$p=0$ is {\it not\/} independent from the other non-zero modes.  In
Fig. \ref{crosscor} we show simulation results for the first term in
the left-hand side of Eq.~(\ref{e1}). Note that the correlation
function saturates asymptotically to a value close to $N/9$.

Let us now remark here that the $t^{1/4}$ behavior of the middle
repton is a consequence of two separate aspects of the middle
repton's motion: (i) that the middle repton simply moves along the
backbone of the polymer (curvilinear motion). For $t\lesssim1$ the
middle repton moves diffusively along the backbone of the
polymer. For $1\lesssim t\lesssim\tau_R\sim N^2$ its motion
corresponds to that of single-file diffusion, which yields the
curvilinear mean-squared displacement of the middle repton
$\langle\Delta u_{N/2}^2(t)\rangle_{\text c}$, where the
subscript `c' denotes an average for a given initial polymer
configuration, increasing in time as $t^{1/2}$. For $\tau_R\lesssim
t\lesssim\tau_d\sim N^3$ the motion of the middle repton becomes
diffusive along the backbone, and its curvilinear motion becomes
diffusive in time, i.e., increases $\sim t/N$, and beyond $\tau_R$ the
backbone of the polymer is completely renewed in physical space. (ii)
The backbone configuration of the polymer, from one end to the other,
is that of a random walk in physical space. Hence, the real-space
mean-squared displacement of the middle repton $\langle\Delta
r_{N/2}^2(t)\rangle$ averaged over all polymer configurations
equals $\sqrt{\langle \Delta u_{N/2}^2(t)\rangle_{\text{c}}}$; i.e.,
$\langle\Delta r_{N/2}^2(t)\rangle\sim t^{1/4}$ for $1\lesssim
t\lesssim\tau_R$, and then $\sim\sqrt{t/N}$ for $\tau_R\lesssim
t\lesssim\tau_d$.

Given the above discussion, one can relate the mean-squared
displacements of the middle repton and the center-of-mass in the
following manner. First, we ignore the end-effects, i.e., assume that
the motion of all the reptons progress along the polymer's backbone,
which is fixed in space. Secondly, for a given backbone of the polymer
fixed in space we imagine an interior repton $i$ of the polymer at
$t=0$. After time $t$, the repton would have moved to a different
location on the backbone, which was occupied by another repton $j$ at
$t=0$: clearly, $j$ is a function of $i$ and $t$, i.e., $j\equiv
j_i(t)$. In fact, the real-space displacement $\Delta\vec r_i(t)$ for
repton $i$ is then written as, using Eq.~(\ref{eq:y}),
\begin{eqnarray}
\Delta\vec r_i(t)=\sum_{k=i}^{j_i(t)}\vec y_k(0).
\label{e31}
\end{eqnarray}
Thereafter, in terms of the link variables $\{\vec y_k\}$s, the real-space
displacement of the center-of-mass for this polymer in time $t$ is given by 
\begin{eqnarray}
\Delta\vec r_{\text{cm}}(t)=\frac1{(N+1)}\sum_{i=1}^{N+1}\Delta\vec r_i(t)=
  \frac1{(N+1)}\sum_{i=1}^{N+1}\sum_{k=i}^{j_i(t)}\vec y_k(0).
\label{e32}
\end{eqnarray}

Thirdly, as we observe that for $1\lesssim t\lesssim\tau_R$,
$\langle\Delta r^2_i(t)\rangle_{\text c}\sim t^{1/2}$, it becomes
clear that $|j_i(t)-i|\sim t^{1/4}$, meaning that in Eq. (\ref{e32})
each {\it distinct\/} $\vec y_k(0)$ appears $\sim t^{1/4}$ times. With
$\langle\vec y_{k_1}(0)\cdot\vec y_{k_2}(0)\rangle=\delta_{k_1k_2}$,
this implies that $\langle
r^2_{\text{cm}}(t)\rangle\sim\sqrt{t}/N$. Note that this argument can
be used again to show that $\langle r^2_{\text{cm}}(t)\rangle\sim t/N$
for $t\lesssim1$ and also that $\langle r^2_{\text{cm}}(t)\rangle\sim
t/N^2$ for $\tau_R\lesssim t\lesssim\tau_d$. Indeed, as for the pure
time behavior is concerned, this argument shows that for a general
reptation motion of a polymer, if the real-space mean-squared
displacement of the middle repton increases in time as $t^{\alpha}$
for some $\alpha$, then the real-space mean-squared displacement of the
center-of-mass has to increase in time as
$t^{2\alpha}/N$. 

\section{Discussion\label{sec:discussion}}

By extensive simulations we have established scaling properties of the
structural modes for a polymer in the repton model. It turns out that
their autocorrelation functions can be written as a function of the
scaling variable $p/N$, where $p$ is the mode number and $N$ the
length of the chain, and a function $A(t)$ which is shared by all
modes. This latter function causes exponential decay at short times
$t<1$ and long times $t>\tau_R \sim N^2$, and causes an intermediate
regime of stretched exponential decay.

Although the notion of structural modes stems from a freely moving
polymer, held together by harmonic forces, it also turns out to be
useful for the dynamics in the repton model, for which the chain is
restricted to move along its contour. The reasons are, apart from the
above mentioned scaling properties, the following.
\begin{enumerate}
\item The structural modes form a complete basis for the structural
  quantities, i.e. any linear function of the link variables can be
  written as a superposition of structural modes.
\item The equal-time correlations between the structural modes are
  strictly orthogonal and they preserve this orthogonality to a high
  degree for correlations at different times.  So for practical
  purposes cross-correlations between structural modes may be ignored.
\item This orthogonality allows to write the correlations in time
  between structural quantities as a sum over the autocorrelation
  functions of the structural modes.
\item The motion of the center-of-mass is correlated with that of the
  middle monomer. This stems from the fact that a reptating polymer
  experiences more than only internal forces --- part of the forces a
  reptating polymer experiences originate from the dynamical
  constraints that prevent the polymer from moving sideways. The
  effect of this correlation is captured in the phenomenon that mode
  $p=0$ is {\it not\/} independent from the other non-zero modes.
\end{enumerate}

The mean-squared displacement of the center-of-mass shows also changes
in its behavior at the same crossover times $t\sim 1$ and $t=\tau_R
\sim N^2$: from an initially diffusive regime in which it grows as
$t/N$ similar to a free Rouse chain, via an intermediate regime with
anomalous diffusion in which it grows as $\sqrt{t/N}$, to an
asymptotic regime with reentrant diffusive behavior where it grows as
$t/N^2$.

The asymptotic regime of the mean-squared displacement of the middle
repton equals that of the center-of-mass, as it should to preserve
integrity of the chain, but this regime is only entered after the tube
renewal time $\tau_d \sim N^3$. It has an initially diffusive regime
in which it grows linearly with time, also similar to the free Rouse
chain. In contrast to the center-of-mass, it shows however two
intermediate regimes in which it grows as $t^{1/4}$ and $\sqrt{t/N}$,
respectively.  We have presented arguments that for general reptation
motion of a polymer, a real-space mean-squared displacement of the
middle repton scaling as $t^\alpha$ for some $\alpha$ must be
accompanied by a mean-squared displacement of the center of mass
increasing in time as $t^{2\alpha}/N$.

Finally we note that the most interesting and revealing aspect of
the mode analysis presented in this paper is the interplay between the
translational degree of freedom and the the structural degrees of
freedom, i.e., the cross-correlation between the position of the
middle repton and the vector connecting the center-of-mass and the
middle repton.

\end{document}